# Log-periodic oscillations due to discrete effects in complex networks

Julian Sienkiewicz, Piotr Fronczak and Janusz A. Hołyst
*Faculty of Physics and Center of Excellence for Complex Systems Research, Warsaw University of Technology, Koszykowa 75, PL-00-662 Warsaw, Poland*

We show that discretization of internode distribution in complex networks affects internode distances $\langle l_{ij} \rangle$ calculated as a function of degrees $k_i k_j$ and an average path length $\langle l \rangle$ as function of network size $N$. For dense networks there are log-periodic oscillations of above quantities. We present real-world examples of such a behavior as well as we derive analytical expressions and compare them to numerical simulations. We consider a simple case of network optimization problem, arguing that discrete effects can lead to a nontrivial solution.

During the last few years much attention has been drawn to average path length issues in complex networks. Several authors [1, 2, 3, 4, 5] have dealt with this problem using different approaches to obtain analytical expressions for average path lengths. One finds a good reason to explore this quantity knowing that it was a small value of the average path length in such systems as social [6] and technological [7] networks that made scientists get interested in this field. Average path length can have different aspects, its value may be just a "chemical distance" between routers or WWW pages [8] but it also appears as "degree of separation" in acquaintances between people [9], number of changes in public transport systems [10, 11] or information handling in a city [12].

In this Letter we study a simple effect of log-periodic oscillations in average path lengths which we observe in several real-world examples. Using a formalism developed in [5] we give a theoretical explanation of this feature supported by numerical simulations of scale-free networks with different scaling exponents. We show that such oscillations are due to discrete effects of path length distributions for networks with large average degree values. We also study a fundamental and well known problem of optimal network density taking into account the shortest average path length and the smallest number of links in a network [13]. We find that the oscillations substantially influence the solution of this problem.

Lately it has been shown [14, 15] that the average distance $\langle l_{ij} \rangle$ between nodes $i$ and $j$ characterized by degrees $k_i$ and $k_j$ can be expressed as:

$$\langle l_{ij} \rangle = a - b \log(k_i k_j). \tag{1}$$

This relation is fulfilled in wide spectrum of real-world networks and their models such as random graphs or Barabási-Albert evolving networks [15], however our recent research shows deviations from this scaling law which take a form of regular oscillations. This can be clearly seen at Fig. 1 where four real-world networks and two common known models have been gathered.

To explain differences between Eq. (1) and plots at Fig. 1 we will use and modify results obtained by Fronczak *et al.* in [5]. In the cited paper exact expressions for average path length using hidden variables formalism have been received. Assuming that each node $i$ is charac-terized by its hidden variable $h_i$ randomly drawn from a given distribution $\rho(h)$ and a connection probability between any pair of nodes is proportional to $h_i h_j$ one can show [17] that a degree distribution $P(k)$ is:

$$P(k) = \sum_h \frac{e^{-h}h^k}{k!}\rho(h). \qquad (2)$$

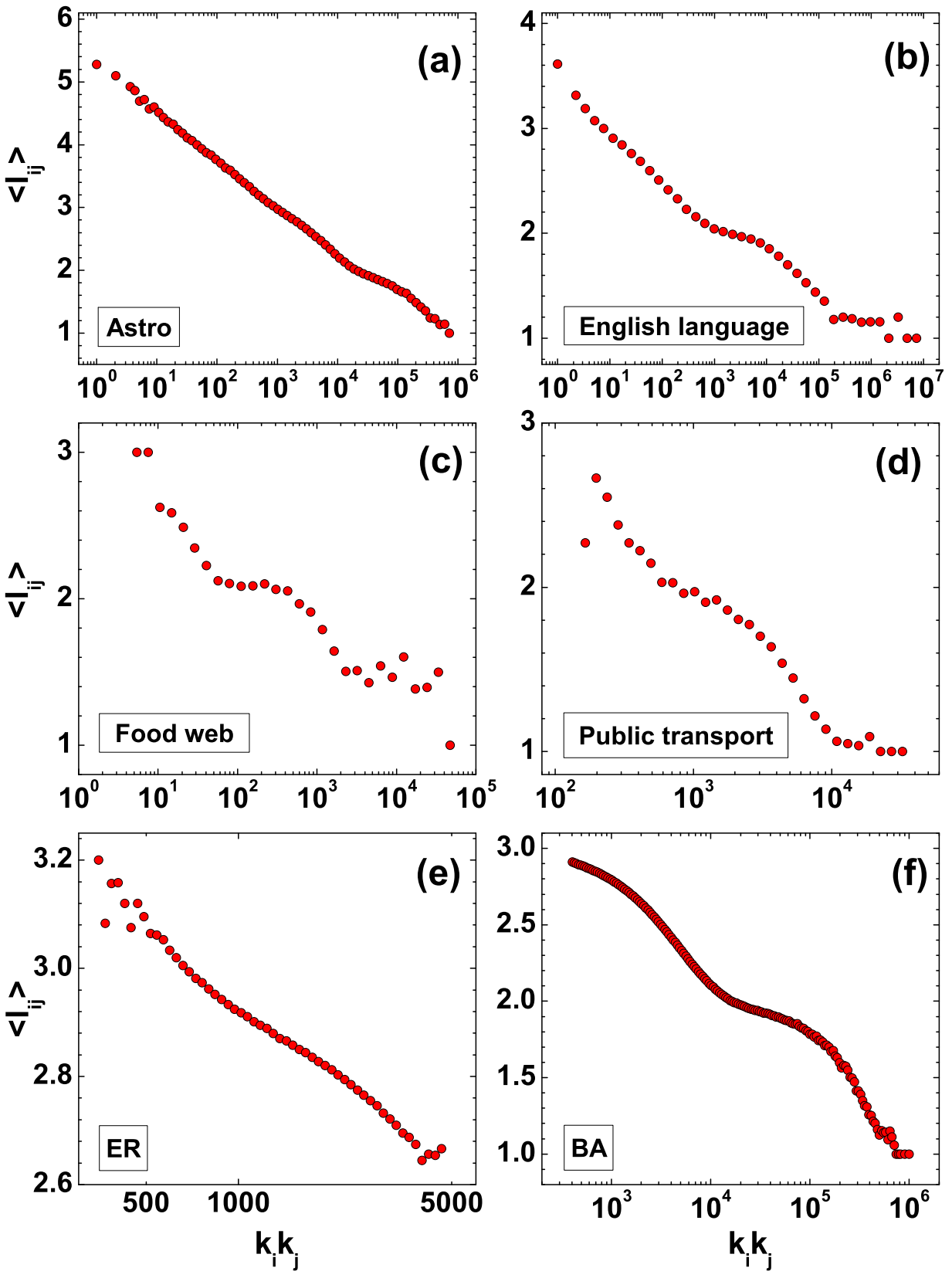

FIG. 1: (color online) Mean distance $\langle l_{ij} \rangle$ between pairs of nodes $i$ and $j$ as a function of a product of their degrees $k_i k_j$ for 4 real-world networks and 2 models. (a) Astro coauthorship network: $N = 13986$ $\langle k \rangle = 25.56$, (b) English language word cooccurrence network $N = 7381$ $\langle k \rangle = 11.98$, (c) Caribbean food web network $N = 249$ $\langle k \rangle = 25.73$, (d) Opole public transport network $N = 205$ $\langle k \rangle = 50.19$. (e) Erdös-Rényi random graph $N = 10000$, $\langle k \rangle = 40$. (f) Barabási-Albert network $N = 10000$ $m = 20$. All data are logarithmically binned. For data sources see [16].

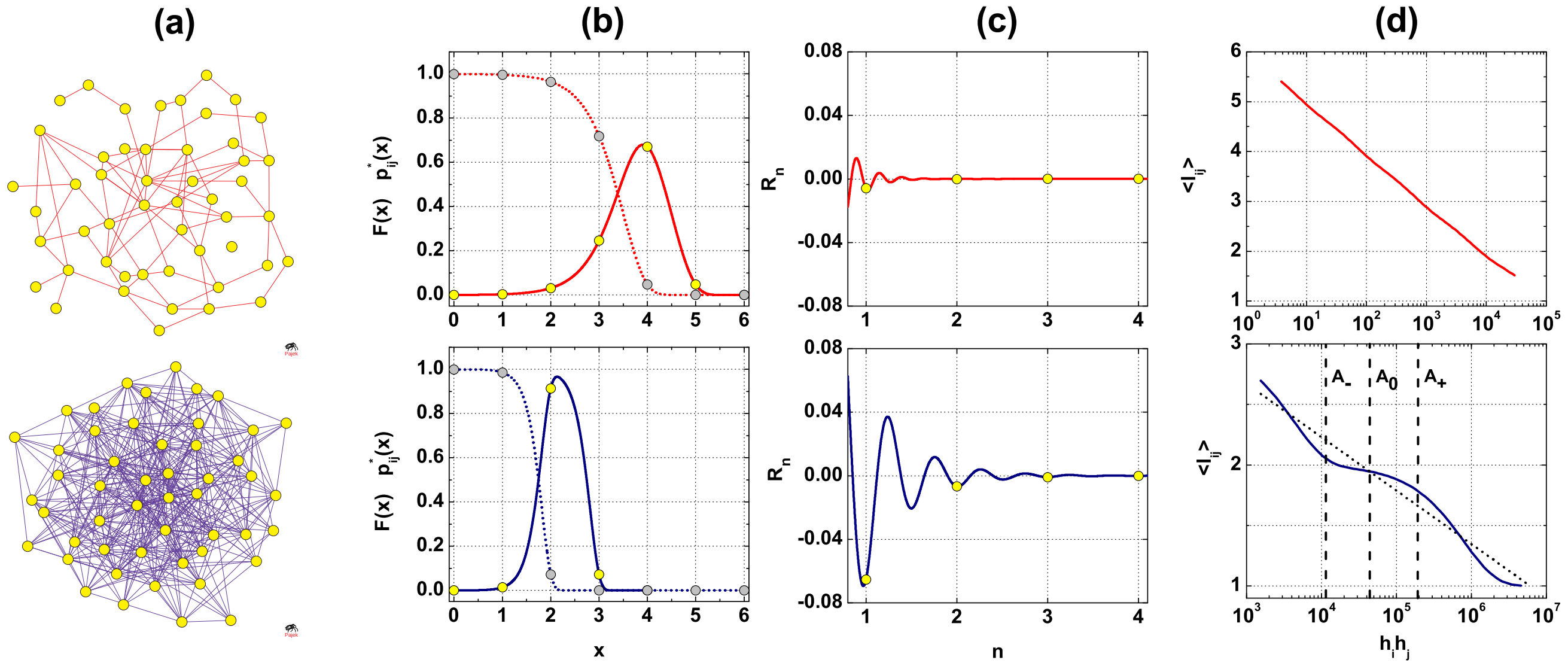

FIG. 2: (color online) Comparison of two networks characterized by hidden variable distribution $\rho(h) = (\alpha-1)m^{\alpha-1}h^{-\alpha}$ for $\alpha = 3.0$ and $N = 10000$ - upper row $m = 2$, lower row $m = 40$. (a) Samples of sparse (upper) and dense (lower) networks, (b, c, d) - detailed description in text and in caption of Figure 3. In case of plots (b) and (c) values of $A$ have been chosen in such a way that the deviation is maximal.

The probability $p^*_{ij}(x)$ that vertices $i$ and $j$ are $x$-th neighbors can be expressed [5] as $p^*_{ij}(x) = F(x-1) - F(x)$, where

$$F(x) = \exp\left(-AB^x\right) \qquad (3)$$

and $A = \frac{h_i h_j}{\langle h^2\rangle N}$, $B = \frac{\langle h^2\rangle}{\langle h\rangle}$. One should have in mind that the parameter $B$ is a "global" one (i.e. its value is determined only by the first and second moment of hidden variable distribution), while $A$ can be called "local" - it depends on a specific product $h_i h_j$. As the expectation value of average distance between $i$ and $j$ can expressed as $\langle l_{ij}\rangle = \sum_{x=1}^{x=\infty} x p^*_{ij}(x) = \sum_{x=0}^{x=\infty} F(x)$, one can write the following equation using Poisson summation formula

$$\langle l_{ij}\rangle = \frac{-\ln A - \gamma}{\ln B} + \frac{1}{2} + R \qquad (4)$$

$$R = \sum_{n=1}^{\infty} R_n \equiv 2\sum_{n=1}^{\infty}\left(\int_0^{\infty} F(x)\cos(2n\pi x)dx\right),$$

where $\gamma = 0.5772$ is Euler's constant. If the average number of links is relatively small then, due to the generalized mean value theorem, the term $R$ can be neglected. Otherwise one must take into account at least the first term from the infinite series in Eq. (4) what leads to log-periodic oscillation $\langle l_{ij}\rangle$ with the period $\Delta\ln(h_i h_j) = \ln B$ (see dicussion below). Figure 2 shows a comparison of such oscillations in sparse ($m = 2$, up-per row) and dense ($m = 40$, lower row) scale-free networks characterized by a hidden variable distribution $\rho(h) = (\alpha - 1)m^{\alpha-1}h^{-\alpha}$ with $\alpha = 3$. The networks have been generated following the procedure $C$ in [18] and represent the class of random networks with asymptotic scale-free connectivity distributions characterized by an arbitrary scaling exponent $\alpha > 2$. At Fig. 2b $F(x)$ (dotted line) and $p^*_{ij}$ (solid line) are presented together with points corresponding to discrete values of those functions. It is clearly seen that for $m = 40$ probability $p^*_{ij}$ is much more narrow than for $m = 2$, thus the slope of $F(x)$ decays more rapidly. Figure 2c shows the cosine transform of $F(x)$ given by the integral in Eq. (4). Depending on the shape of $F(x)$, the amplitude of this transform can take small/large values resulting in small/large values of $R$. One should keep in mind that because $R$ is in fact a sum of discrete values of a given transform taking only the first term in the sum (i.e. $n = 1$) is sufficient to obtain well approximated value of $R$ (cf. points corresponding to discrete values of $R_n$ at Fig. 2c). Figure 2d shows resulting average distance $\langle l_{ij}\rangle$ between nodes $i$ and $j$ as a function of hidden variables $h_ih_j$ without (dotted lines) and with (solid lines) included term $R$. In case of sparse network the $R$ term can be omitted (curves overlap), while for a dense one its value modifies the shape of $\langle l_{ij}\rangle$ a lot.

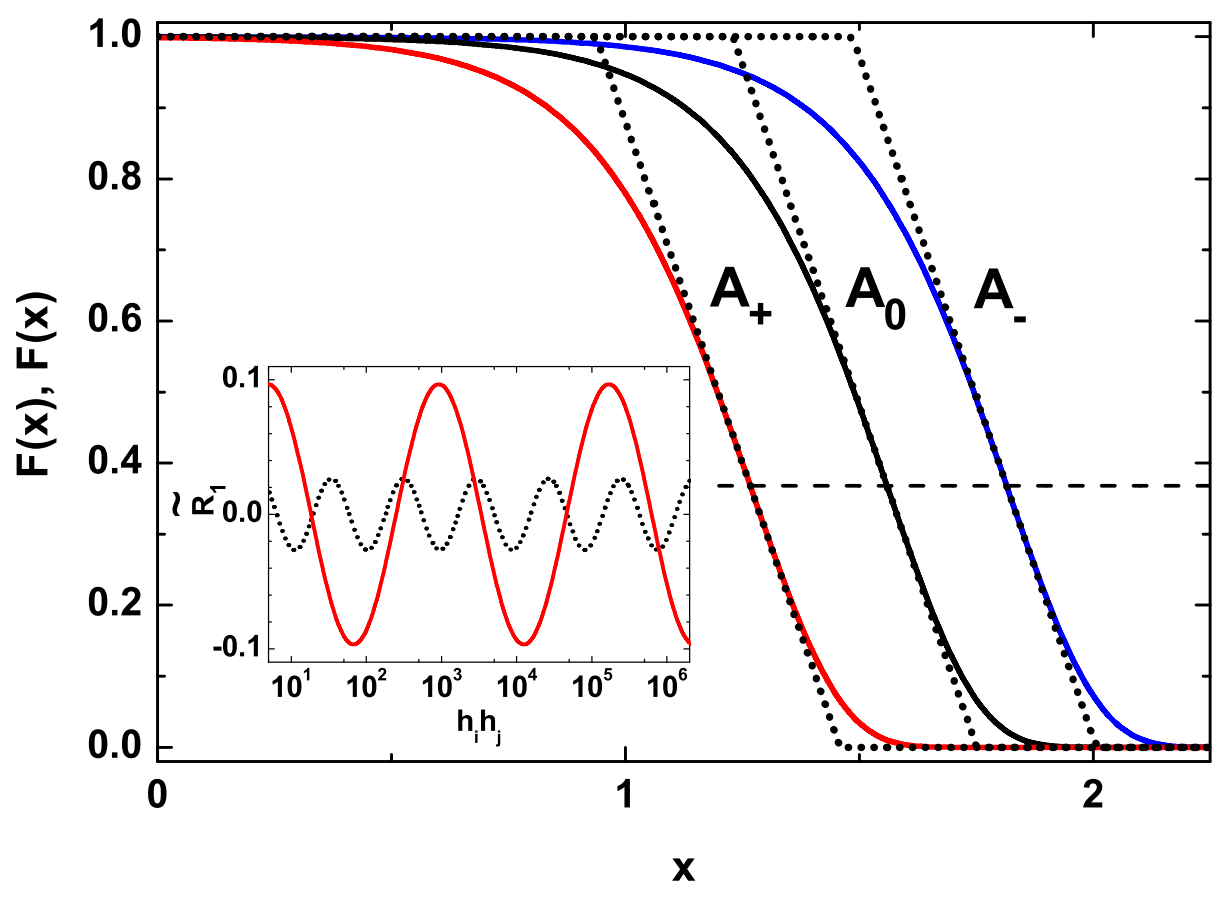

FIG. 3: (color online) Function $F(x)$ (solid lines) and its linear approximation $\widetilde{F}(x)$ (dotted lines) for scale-free network with $\alpha = 3$, $N = 10000$ and $m = 40$ calculated for three different values of product $h_i h_j$ (see labelled dashed lines at Fig. 2): ($A_-$) $h_i h_j = 11389$ - maximal negative deviation from $\langle l_{ij}\rangle$ trend ($A_0$) $h_i h_j = 43249$ - minimal (zero) deviation from $\langle l_{ij}\rangle$ trend ($A_+$) $h_i h_j = 198730$ - maximal positive deviation from $\langle l_{ij}\rangle$ trend. Dashed line represents the point of inflexion $x_i$ of $F(x)$ ($F(x_i) = 1/e$) used to calculate tangent of $F(x)$. Inset shows $\widetilde{R}_1$ versus product $h_i h_j$ in case of $m = 2$ (dotted line) and $m = 40$ (solid line).

To obtain more quantitative results one should perform the integral in Eq. (4), however it is not analytical, so in order to calculate the term $R$ one can approximate $F(x)$ with the following piecewise linear function $\widetilde{F}(x)$

$$\widetilde{F}(x) = \begin{cases} 1 & x < x_0, \\ \frac{1}{e}\left(1 - \ln A - x\ln B\right) & x \in < x_0, x_1 >, \\ 0 & x > x_1, \end{cases} \quad (5)$$

where $x_0 = (1 - \ln A - e)/\ln B$ and $x_1 = (1 - \ln A)/\ln B$. Since the function $F(x)$ is translationally invariant with respect to the argument $x$ after rescaling the parameter $A$ ($F(x; A) = F(x - x'; A')$) one can freely choose the point in which the slope coefficient is calculated as the tangent of $F(x)$. In order to simplify the calculation we have chosen the inflexion point $x_i$ of $F(x)$. Functions $\widetilde{F}(x)$ and $F(x)$ are presented at Fig. 3. Using Eq. (5) one can approximate terms $R_n$ with

$$\widetilde{R}_n = -\frac{\ln B \sin\left(\frac{\pi n e}{\ln B}\right)}{\pi^2 n^2 e} \sin\left[\frac{\pi n}{\ln B}\left(2\ln A - 2 + e\right)\right]. \quad (6)$$

As one can see taking only the first term in the above series is justified because next terms decay as $1/n^2$. Equation (6) allows us to make an immediate observation that deviations from Eq. (1) take the form of regular oscillations along $h_ih_j$ axis with period equal to $\ln B$ which increases with the heterogeneity of the networks (see inset at Fig. 3). This very value is forced by the *discrete nature of distance* in network - the period along $\langle l_{ij}\rangle$ is equal to 1 and the tangent of the function $\langle l_{ij}\rangle(h_ih_j)$ is $(\ln B)^{-1}$ (see Eq. (4)). One can also easily calculate that the deviation vanishes as long as $\langle l_{ij}\rangle \approx k/2$ where $k$ is an integer. For dense networks the amplitude of oscillations grows monotonically with $B$ - that is why the effect of oscillations is visible only in sufficiently dense networks. Figure 4 presents a comparison of average distance $\langle l_{ij}\rangle$ versus $h_ih_j$ for scale-free networks with different scaling exponents $\alpha$. As expected, the amplitude of oscillations rises with decaying $\alpha$, which can be easily understood as $\ln B_{\alpha_1} > \ln B_{\alpha_2}$ for $\alpha_1 < \alpha_2$. Similar oscillations effects can also be observed for average path length $\langle l\rangle$, which value is obtained by integration of Eq. (4) over all pairs of products $h_ih_j$ (see inset (b) at Fig. 5).

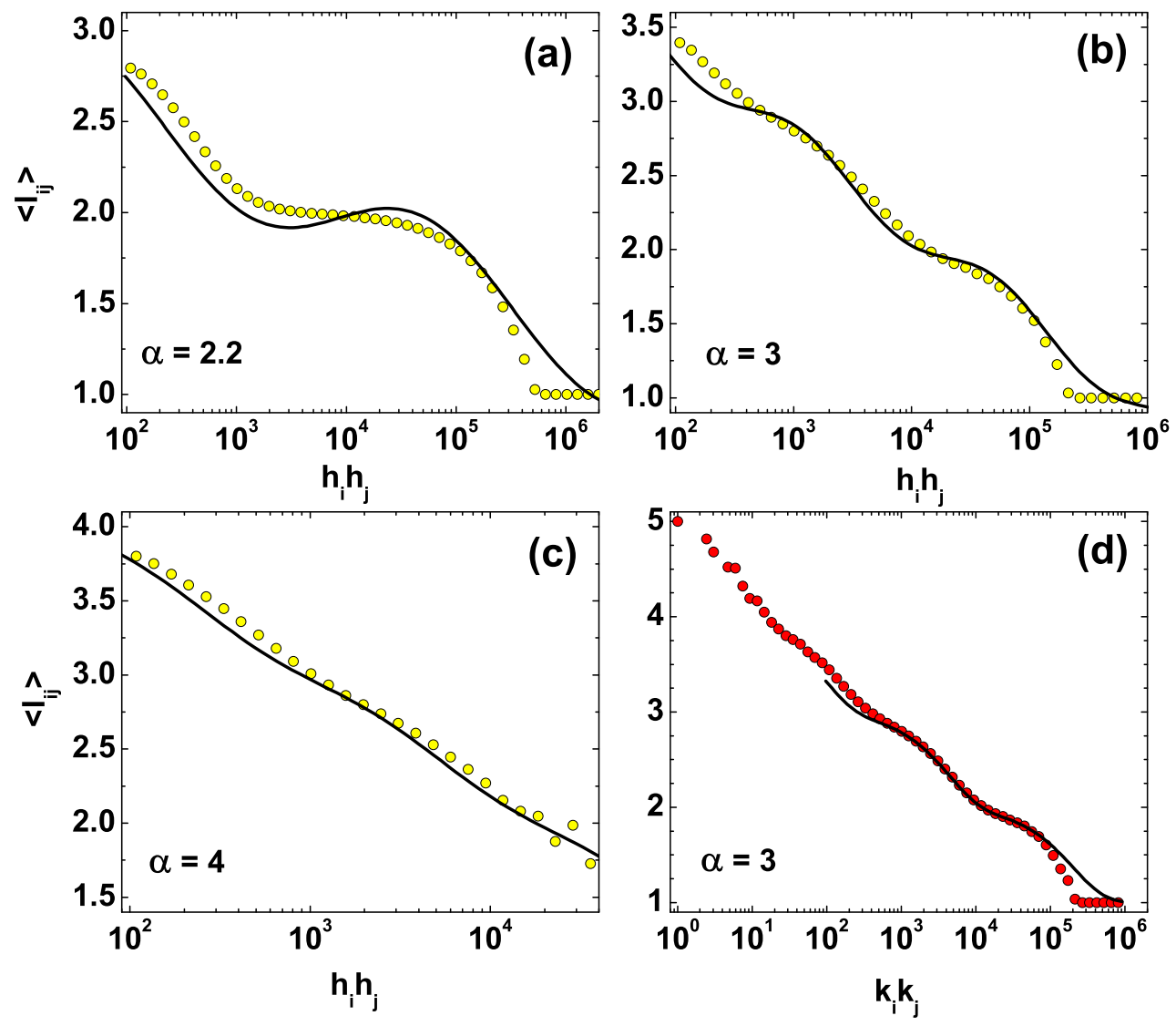

FIG. 4: (color online) Average distance $\langle l_{ij}\rangle$ between nodes $i$ and $j$ versus their hidden variable product $h_ih_j$ (plots (a), (b) and (c)) or $k_ik_j$ (d) for scale-free networks of $N = 10000$ nodes and $\alpha = 2.2$ (a), $\alpha = 3$ (b) and (d) and $\alpha = 4$ (c). Scatter data are obtained using algorithm presented in [18] while solid lines have been calculated from Eq. (4) where $R$ is taken directly from Eq. (6).

Let us now focus on possible applications of the presented phenomenon. One of them can be a network optimization process which has been widely studied in recent years [13, 19, 20]. Such an optimization is of common interest in many different areas, among them electrical engineering, telecommunication, road construction and trade logistics. The simplest model is based on the assumption of minimal transport costs. These costs include two main aspects of network performance: a price of constructing and maintaining links between nodes and a price caused by communication delays of information transfer. The former one is proportional to the total number of links (we assume the same price for every link), while the later one should be proportional to the sum of the shortest existing connections between each two nodes:

$$C = (1-\lambda)\,\frac{N}{2}\langle k\rangle + \lambda\binom{N}{2}\langle l\rangle. \tag{7}$$

Here $\lambda$ is a parameter controlling a ratio between prices of a single link and costs of communication delays. In fact one has to find an optimal link density considering two contradictive demands: a fully connected network with the shortest connections and a tree with the smallest number of links. A typical solution of this problem is a unimodal cost function with minimum at some intermediate value of $\langle k\rangle$. Discrete effects in networks studied above can lead to reshaping of the total cost function. As an example let us consider the scale-free network generated by method described in [18] with parameters $N = 10^6$ and $\alpha = 3$. The cost function for this network is presented at Fig. 5 (we also show how this function could look like if we neglected discrete effects). One can see that neglecting of the correction term can lead to about 30% underestimation of optimal network density. Inset (a) at this figure obtained for another value of the parameter $\lambda$ shows different situation - instead of one global minimum we have now two well separated minima. The network administrator who tries to operate in accordance with the economic rule (7) has just to remember that the improvement of network efficiency can lead to a temporal increase of costs and can be discouraging since one has to pass over the cost barrier. Much simpler application of the observed phenomenon is presented at Fig. 5b, where one can see that during the network growth there are regions where average path length increases slower (faster) which can encourage (discourage) the network administrator for further network expansion.

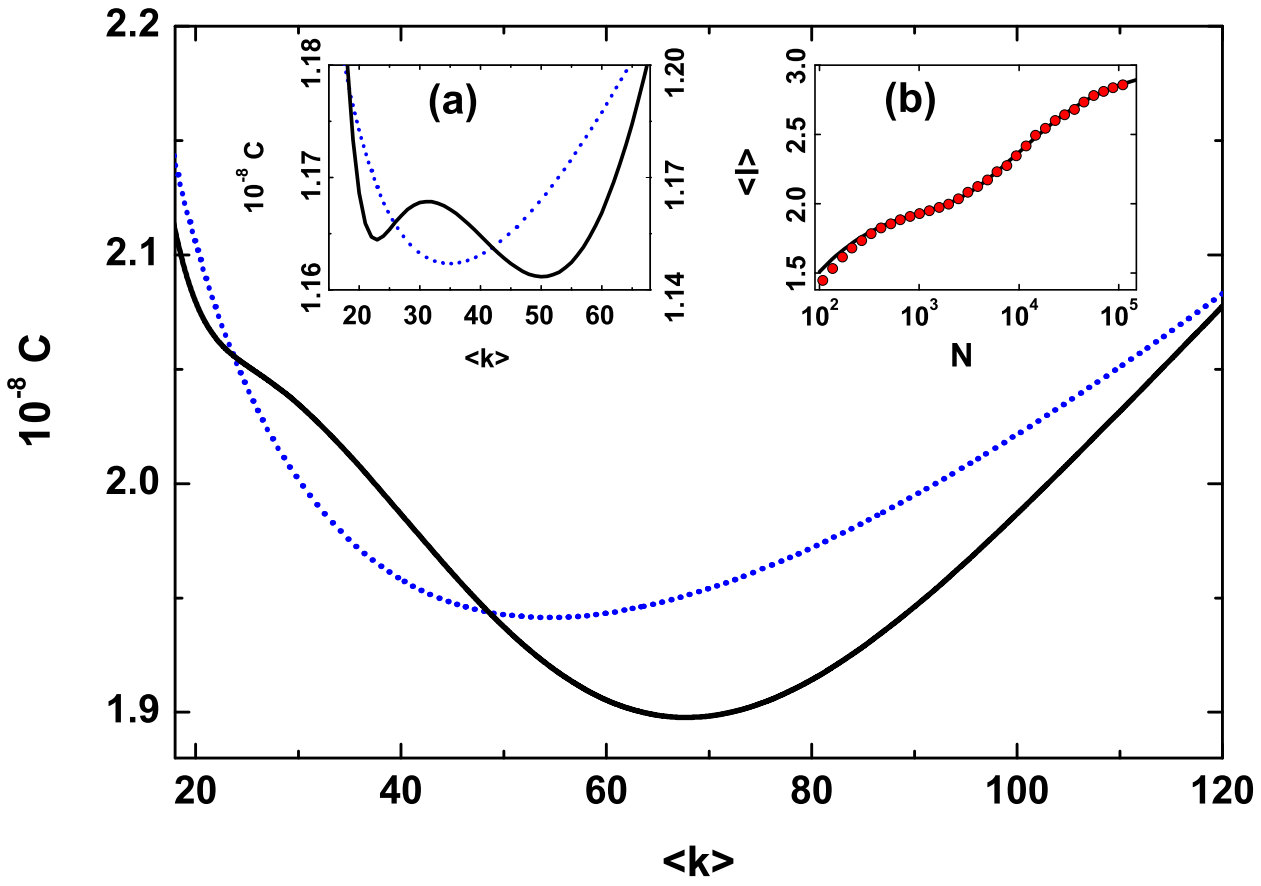

FIG. 5: (color online) Cost function $C$ versus average degree $\langle k\rangle$ for scale-free network characterized by $N = 10^6$ nodes, $\alpha = 3.0$ and $\lambda = 10^{-4}$. Solid line is obtained assuming oscillations' correction while dotted line neglects it. Inset (a) shows cost function $C$ for identical network parameters $N$ and $\alpha$ but with $\lambda = 5.4 \cdot 10^{-4}$. Left Y-axis corresponds to cost function with oscillations' correction (solid line) while right Y-axis corresponds to function that neglects the correction (dotted line). Inset (b) presents average path length $\langle l\rangle$ versus system size $N$ for scale-free network with $\alpha = 3$ and $m = 40$: solid line is theory, while scatter data have been obtained using the hidden variable algorithm [18].

To summarize: we have presented an explanation of the oscillations in relations $\langle l_{ij}\rangle(k_i k_j)$ and $\langle l\rangle(N)$ observed in real-world networks starting from scientific collaboration and ending at public transport systems. We have also provided examples of the influence this effect might have for simple optimization problems.

JS and JAH acknowledge a support from the EU Grant *Measuring and Modelling Complex Networks Across Domains* - MMCOMNET (Grant No. FP6-2003-NEST-Path-012999) and from Polish Ministry of Education Science (Grant No. 13/6.PR UE/2005/7). PF acknowledges a support from the EU Grant CREEN FP6-2003-NEST-Path-012864. The authors are thankful to Agata Fronczak for fruitful discussions.

[1] M. E. J. Newman, C. Moore, and D. J. Watts, Phys. Rev. Lett. **84**, 3201 (2000).

[2] G. Szabó, M. Alava, and J. Kertész, Phys. Rev. E **66**, 026101 (2002).

[3] R. Cohen and S. Havlin, Phys. Rev. Lett. **90**, 058701 (2003).

[4] S. N. Dorogovtsev, J. F. F. Mendes, and A. N. Samukhin, Nucl. Phys. B. **653**, 307 (2003).

[5] A. Fronczak, P. Fronczak, and J.A. Hołyst, Phys. Rev. E **70**, 056170 (2004).

[6] D. J. Watts and S. H. Strogatz, Nature (London) **393**, 440 (1998).

[7] R. Albert, H. Jeong, and A.-L. Barabási and R. Albert, Nature (London) **401**, 130 (1999).

[8] R. Pastor-Satorras and A. Vespignani, *Evolution and Structure of the Internet: A Statistical Physics Approach* (Cambridge University Press, Cambridge, 2004).

[9] S. Milgram, Psychology Today **2**, 60 (1967).

[10] P. Sen, S. Dasgupta, A. Chatterjee, P. A. Sreeram, G. Mukherjee, and S. S. Manna, Phys. Rev E **67**, 036106 (2003)

[11] J. Sienkiewicz and J.A. Hołyst, Phys. Rev E **72**, 046127 (2005).

[12] M. Rosvall, A. Trusina, P. Minnhagen, and K. Sneppen, Phys. Rev. Lett. **94**, 028701 (2005).

[13] F. Schweitzer, *Brownian Agents and Active Particles* (Springer, Berlin, 2003).

[14] A.E. Motter, T. Nishikawa, Y.C. Lai, Phys. Rev. E **66** 065103(R) (2002).

[15] J.A. Hołyst, J. Sienkiewicz, A. Fronczak, P. Fronczak, and K.Suchecki, Phys. Rev E **72**, 026108 (2005).

[16] Soures for data are following: data for Astro network have been collected from publicly avaible database at http://arxiv.org, English coocurrance dataset has been downloaded from V. Batagelj WWW page http://vlado.fmf.uni-lj.si/pub/networks/pajek/ and Caribbean food web data have been taken from The Integrative Ecology Group http://ieg.ebd.csic.es. Opole

data have been collected for so called "space P" [11] which is definied as follows: nodes are bus, tram or underground stops and an edge means that there is direct route linking them.

[17] M. Boguñá and R. Pastor-Satorras, Phys. Rev. E **68**, 036112 (2003).

[18] A. Fronczak, P. Fronczak, e-print cond-mat/0503069 (2005).

[19] M. T. Gastner and M. E. J. Newman, Eur. Phys. J. B **49**, 247 (2006).

[20] R. Ferrer i Cancho and R. Solé, *Statistical Physics of Complex Networks, Lecture Notes in Physics* (Springer, Berlin, 2003).